\documentclass[PAPER, atlasdraft=false, UKenglish, cernpreprint, texmf, orcidlogo]{atlasdoc}
\usepackage[subcaption]{atlaspackage}
\usepackage{atlasbiblatex}

\usepackage{atlascontribute}

\usepackage{atlasphysics}
\graphicspath{{logos/}{figures/}}

\usepackage{ANA-STDM-2023-21-PAPER-defs}
\DeclarePairedDelimiter{\abs}{\lvert}{\rvert}
\usepackage{amsmath,bm}
\usepackage{xspace}
\usepackage{multirow}

\AtlasTitle{Combined effective field theory interpretation of measurements sensitive to quartic gauge boson couplings in $pp$ collisions at $\sqrt{s}=13$~\TeV with the ATLAS detector}

\AtlasAbstract{%
A combination of measurements sensitive to anomalous quartic electroweak gauge boson couplings is presented using proton--proton collision data collected by the ATLAS detector at $\sqrt{s} = 13$~\TeV at the LHC\@. Contributing analyses include measurements of vector-boson scattering in numerous final states as well as a tri-boson measurement.  The combined measurement is used to constrain anomalous electroweak boson quartic self-couplings that result from dimension-8 operators in the \'{E}boli model using an effective field theory.  Results are presented as 68\% and 95\% confidence level intervals parameterised by one or two Wilson coefficients, both with and without unitarity constraints applied. Theoretical bounds from unitarity and positivity are overlaid where relevant.  Confidence intervals obtained from simultaneous profiled fits to all Wilson coefficients are also presented.
}

\AtlasRefCode{STDM-2023-21}

\PreprintIdNumber{CERN-EP-2026-076}

\AtlasJournalRef{\PLB 879 (2026) 140683}
\AtlasDOI{DOI:10.1016/j.physletb.2026.140683}

\AtlasJournal{\PLB}

\hypersetup{pdftitle={ATLAS document},pdfauthor={The ATLAS Collaboration}}

\begin{document}

\maketitle

\tableofcontents

%
%

%

% The next lines are included from the .//tex/Introduction.tex input file
\section{Introduction}
\label{sec:intro}

The Large Hadron Collider (LHC) has enabled a rigorous study of the Standard Model (SM) up to the electroweak scale and beyond. Numerous dedicated analyses searching for new \TeV-scale particles have been conducted by both the ATLAS and CMS collaborations, see for example Ref.~\cite{EXOT-2022-38,CMS-B2G-18-006}, without finding any significant excess above the SM background prediction. In the absence of direct evidence of new physics, a complementary program is to search for small deviations from SM predictions that can arise from new physics with energy scales beyond that of direct production at the LHC\@. The SM effective field theory (EFT)~\cite{brivio2019standard} is a theoretical framework that describes potential new physics at or above a scale $\Lambda$ in terms of SM gauge invariant operators of dimension-$d$, which are suppressed by the scale $\frac{E^{d-4}}{\Lambda^{d-4}}$, where $E$ is the energy scale of the process. Wilson coefficients, denoted $f_i$, parameterise the size of each potential operator contribution.

Constraints on dimension-6 operators were extracted from combinations of results from Higgs boson, electroweak, top-quark, and jet related measurements using 13~\TeV LHC data from both the ATLAS and CMS collaborations~\cite{HIGG-2022-17,CMS-SMP-24-003}, and electroweak precision observables at the $Z$ mass performed at LEP and SLC~\cite{electroweak2006precision}.
Such results are truncated at dimension-6 since these are the leading-order new-physics contributions to most processes, excluding odd-dimension operators that violate baryon and lepton number conservation.
Many potential beyond the SM scenarios predict direct anomalous Quartic-Gauge Couplings (\aQGCs)~\cite{eboli2004bosonic,eboli2006pp}. Others predict resonant new physics beyond the \TeV scale that could be experimentally observed at current collider experiments as \aQGCs, for example composite Higgs boson models~\cite{contino2011effect}, warped extra-dimension models~\cite{randall1999large}, gravitons~\cite{agashe2007warped}, two-Higgs-doublet models~\cite{branco2012theory} and unified models~\cite{georgi1974unity}.
While dimension-6 operators can indirectly affect processes sensitive to \aQGCs through diagrams with triple-gauge boson vertices, operators affecting only quartic-gauge couplings begin to appear at dimension-8.
Thus, such models are potentially not well constrained by dimension-6 results.

\begin{figure}[h!]
\centering
\subfloat[]{\label{fig:feynmana}\includegraphics[height=3.3cm]{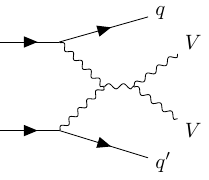}}
\subfloat[]{\label{fig:feynmanb}\includegraphics[height=3.3cm]{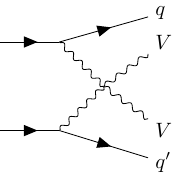}}
\subfloat[]{\label{fig:feynmanc}\includegraphics[height=3.3cm]{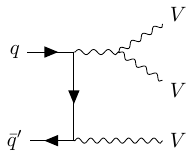}}
\subfloat[]{\label{fig:feynmand}\includegraphics[height=3.3cm]{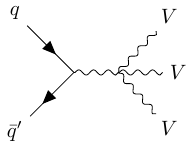}}
\caption{Representative Feynman diagrams for a) vector-boson scattering with triple-gauge vertices, b) vector-boson scattering with quartic-gauge vertices, c) tri-boson production with triple-gauge vertices, and d) tri-boson production with quartic-gauge vertices. The \aQGCs act only on diagrams b) and d) through anomalous values or vertices forbidden in the SM.\label{fig:feynman}}
\end{figure}

At dimension-8 there are tens of thousands of new operators along with multiple insertions of dimension-6 operators, making probing all operators experimentally impractical.
The effects of dimension-8 operators that give rise to \aQGCs are most experimentally prevalent in the vector-boson-scattering (VBS) process and tri-boson production.
Representative Feynman diagrams for the VBS and tri-boson processes are shown in Figure~\ref{fig:feynman}.
The results presented here use the \Eboli model~\cite{eboli2006pp} as the basis for the dimension-8 operators that predict \aQGCs. The dimension-6 operators that cause anomalous triple gauge couplings that would contribute to the analyzed VBS and tri-boson channels (Figure~\ref{fig:feynmana} and \ref{fig:feynmanc}) are assumed to be sufficiently well constrained by dedicated searches such that their contributions are negligible.

Measurements of VBS production have been reported by the ATLAS and CMS
collaborations using the full Run~2 13~\tev data in numerous experimental channels~\cite{STDM-2018-32,CMS-SMP-17-004,STDM-2022-06,CMS-SMP-21-001,STDM-2018-35,CMS-SMP-18-001,STDM-2020-02, CMS-SMP-20-001,STDM-2018-31, CMS-SMP-21-011,STDM-2018-59,STDM-2018-36,CMS-SMP-20-016,STDM-2018-27,CMS-SMP-20-013}.
Similarly, tri-boson production has been measured in several experimental channels by the ATLAS~\cite{STDM-2018-33,STDM-2019-09,STDM-2020-08} and CMS~\cite{CMS-SMP-19-014,CMS-SMP-19-013,CMS-SMP-22-006,CMS-SMP-22-018} collaborations.

This Letter presents a combination of measurements from seven VBS analyses ~\cite{STDM-2018-32,STDM-2018-35,STDM-2020-02,STDM-2018-31,STDM-2018-59,STDM-2018-36,STDM-2018-27} and one tri-boson analysis~\cite{STDM-2018-33}, each performed using 140~\ifb of pp collision data recorded by the ATLAS detector during Run~2 of the LHC\@. A combined likelihood is built from the likelihoods of the input measurements and a likelihood fit is performed to extract constraints on dimension-8 Wilson coefficients. Constraints on the Wilson coefficients of 17 of the \'{E}boli basis operators are presented when considering one or two operators at a time, with and without unitarity constraints.
Profiled confidence intervals obtained from simultaneous fits to the 17 operators are also presented.

% End of text imported from the .//tex/Introduction.tex input file

%

% The next lines are included from the .//tex/Theory.tex input file
\section{Theoretical framework}
\label{sec:theory}

The SM Effective Field Theory provides a model-independent description of effects beyond the SM that introduce new particles at a mass scale $\Lambda$, which is large in comparison to the electroweak scale.
At energies significantly below the scale $\Lambda$, the effect of new physics can be described by additional operators ${\mathcal{O}_i^{(d)}}$ of energy dimension $d>4$ that extend the SM Lagrangian in orders of $E/\Lambda$, where $E$ is the energy exchanged in the process.
These operators are built from combinations of SM fields that obey the symmetries of the SM such as gauge invariance.
Ignoring odd-dimensional operators, which introduce lepton and baryon number violation, and dimension-6 terms, which are highly constrained by existing results, the leading terms in this effective Lagrangian become
\begin{equation}
\mathcal{L}_\textrm{Eff} = \mathcal{L}_\textrm{SM}^{(4)} + \sum\limits_i \frac{f_i^{(8)}}{\Lambda^4} \mathcal{O}_i^{(8)} %
\label{eq:smeft-dim8}
\end{equation}
where $\mathcal{L}^{(4)}_{SM}$ is the SM Lagrangian, which contains dimension-2 and dimension-4 operators. The remaining terms $\mathcal{O}_i^{(8)}$ denote the new dimension-8 operators, and $f_i^{(8)}$ are the corresponding dimensionless Wilson coefficients.

With the Lagrangian given in Equation \ref{eq:smeft-dim8}, scattering cross-sections can be approximated as
\begin{equation}
\begin{aligned}
\sigma
&= \abs{\mathcal{M}_\textrm{SM}}^2 + \sum_i \frac{f^{(8)}_i}{\Lambda^4}2\mathrm{Re}\left({\mathcal{M}^{(8)}_i\mathcal{M}^*_\textrm{SM}}\right) + \sum_i \frac{\left(f^{(8)}_i\right)^2}{\Lambda^8}\abs{\mathcal{M}^{(8)}_i}^2 + \sum_{i<j} \frac{f^{(8)}_if^{(8)}_j}{\Lambda^8}2\mathrm{Re}\left({\mathcal{M}^{(8)}_i\mathcal{M}^{(8)*}_j}\right) %
\label{eq:smeft-operator-expansion}
\end{aligned}
\end{equation}
where $\mathcal{M}_\textrm{SM}$ is the matrix element as it occurs in the SM, and $\mathcal{M}^{(8)}_{i}$ are the new matrix elements calculated from the operators $\mathcal{O}^{(8)}_i$.
The first and third terms refer to the SM and the dimension-8 squared amplitude contributions to the cross-section, respectively. The second term, which is of order $\Lambda^{-4}$, is the interference between the SM and the dimension-8 operators.
Due to the scaling behavior of the SM-interference and the dimension-8 squared amplitude contributions to the cross-section relative to the Wilson coefficients, the second and third terms are referred to as the linear and quadratic contributions of the expansion, respectively.
The last term represents the interference of dimension-8 operators with one another, and is referred to as the cross-term.

In this work, confidence intervals on dimension-8 operator couplings are set using the \Eboli model~\cite{eboli2006pp}, which provides a subset of the total dimension-8 operators satisfying the SM $SU(2)_L \times U(1)_Y$ symmetry and specifically producing \aQGCs.
In this model, the 21 new C-even P-even dimension-8 operators can be categorized into: scalar-types that only contain derivatives of the scalar Higgs field ($O_{S0}$, $O_{S1}$, $O_{S2}$), tensor-types that only contain electroweak boson field strengths ($O_{T0}$, ..., $O_{T9}$), and mixed-type operators ($O_{M0}$, ..., $O_{M7}$).
The $O_{M6}$ term was found to be redundant in Ref.~\cite{eboli2016mapping} and is not considered in this work.
Similarly, the $O_{T3}$ and $O_{T4}$ operators are excluded due to the fact this \textsc{MadGraph5\_aMC@NLO} simulation implementation was not available in many of the individual analysis publications.
Following Ref.~\cite{STDM-2018-35,STDM-2018-32,STDM-2018-27} the operators $O_{S0}$ and $O_{S2}$ are varied simultaneously with equal coefficient values $f_{S02}$.
This leaves 17 independent Wilson coefficients, all of which are considered in this work.
It is worth noting that due to the helicity structure of the \Eboli operators, the SM-interference term in Equation~\ref{eq:smeft-operator-expansion} is highly suppressed in the \Eboli basis.

Theoretical constraints limit the regime of validity of the EFT operator expansion, and are considered in the result. Operators with mass dimension greater than four can lead to interaction probabilities that increase until unitarity is violated. Theoretical bounds are thus obtained by requiring that perturbative partial-wave unitarity is satisfied as derived in Ref~\cite{eboli2020,eboli2026unitarity}.
In addition, assuming the new high-energy physics obeys the low-scale principles of unitarity, locality, and causality, cross-sections on the helicity-amplitude must be positive definite at all scales.
Positivity bounds are computed for the \Eboli dimension-8 basis in Ref~\cite{positivity_bounds}.

Simulated samples of VBS and tri-boson production under the effect of the \Eboli model were simulated using \textsc{MadGraph5\_aMC@NLO}~(v3.5.5)~\cite{Alwall:2014hca} at leading-order in quantum chromo-dynamcis (QCD) with the NNPDF30\_nlo\_as\_0119 parton distribution function set~\cite{Ball:2014uwa}.
Decays of the vector bosons were handled with \textsc{MadSpin}~\cite{Artoisenet:2012st}. Parton showering, hadronisation, and underlying event modeling were performed using \textsc{Pythia~8} with the %
dipole recoil scheme enabled~\cite{Sjostrand:2014zea}.
The total cross-section in Equation~\ref{eq:smeft-operator-expansion} was decomposed, with the linear, quadratic, and cross-term EFT contributions simulated separately.
The SM component of the expansion uses the original simulated samples from the individual analyses that may be at higher-orders in QCD\@.
The total process including the EFT effects for a given Wilson coefficient value is then reproduced by adding the EFT contributions, scaled appropriately by the Wilson coefficients, to the predicted SM contribution.

% End of text imported from the .//tex/Theory.tex input file

% The next lines are included from the .//tex/Inputs.tex input file
\section{Input measurements}
\label{sec:inputs}

All measurements included in the combination were performed using 140 fb$^{-1}$ of $\sqrt{s}=13$ \TeV $pp$ collision data recorded with the ATLAS detector at the LHC\@. The input measurements used to constrain \aQGC parameters are summarized in Table~\ref{tab:inputs}.
The individual measurements provide either reconstruction-level (reco-level) results in a differential distribution, unfolded results of a differential distribution, or a reconstruction-level inclusive cross-section.
Several analyses were published with an earlier measured integrated luminosity value of 139 fb$^{-1}$ with a larger uncertainty. For this result, these inputs were adjusted to reflect the more precise, updated measured values.

\begin{table}
\caption{Summary of input analyses in the combination. Full, Partial, and None refer to whether all relevant operators, a subset of them, or no dedicated interpretation were explored in the original publication.}
\resizebox{\textwidth}{!}{\begin{tabular}{llccc}
\toprule
Analysis & Input type & \'{E}boli operator results & Luminosity [\ifb] & Reference \\
\midrule
\VVSemi   & Reco-level Differential & Full & 140 & \cite{STDM-2018-27} \\
\WZ       & Reco-level Differential & Partial & 140 & \cite{STDM-2018-35} \\
\ssWW     & Reco-level Differential & Full & 139 & \cite{STDM-2018-32} \\
\ZZllll   & Unfolded Differential   & None & 140 & \cite{STDM-2020-02} \\
\Wy       & Unfolded Differential   & Full & 140 & \cite{STDM-2018-31} \\
\Zvvy     & Reco-level Inclusive    & Partial & 139 & \cite{STDM-2018-59} \\
\Zlly     & Unfolded Differential   & None & 140 & \cite{STDM-2018-36} \\
\Wyy      & Reco-level Differential & None & 140 & \cite{STDM-2018-33} \\
\bottomrule
\end{tabular}}
\label{tab:inputs}
\end{table}

To facilitate the statistical interpretation, it is preferable that each included analysis region has no statistical overlap so that each data event is only counted once in the combined result.
There is a non-negligible overlap found between the control region of the \ssWWShort measurement that is used to determine the shape of the \WZShort background, and the signal region of the \WZShort measurement. This overlapping control region is dropped from the combination to avoid double counting events. This choice has a negligible effect on the results.
The overlap between analyses is at the sub-percent level in all other cases.

Several adjustments are made to the input results to facilitate full coverage of the operator parameter space across experimental channels.
In the \ZvvyShort \aQGC results, only a single overflow bin in leading photon \pt was used in the original EFT interpretation, with a threshold value varying as function of the operator and of the clipping value. For this analysis the full leading photon \pt distribution is used in eight bins, as it facilitates the combination with the other analyses and allows probing several operators at the same time. This change improves the confidence intervals by 1\%.
For \WyyShort, the reco-level result was used, splitting the final observable of leading photon \pt into three bins $[25, 45, 80, \infty]$~\GeV, as opposed to the single-bin result used in the publication. The data-driven background estimation procedure was reproduced and validated, as was done in the original publication, with this finer binning. This finer binning improves the constraints by up to 40\%.
In its original publication, the \WyShort result used a different kinematic observable for each operator under consideration. For simplicity in this combination, the unfolded \pt of the leading lepton distribution is used as the final observable, which at most degrades the confidence intervals on a subset of operators by 20\%.

In several of the individual measurement publications, either none or only a subset of the operators were probed.
For completeness, this result considers the potential impact of all operators in all channels.
The published \ZllyShort, \ZZllllShort  and \WyyShort measurements do not provide a dedicated \Eboli model result, and so in this Letter a dedicated interpretation is conducted for each of these analyses.
For the \ZllyShort (\ZZllllShort) result, the publicly available one(two)-dimensional unfolded distribution of the \pt of the $Z\gamma$ system (invariant masses of the $ZZ$ and dijet systems), are reinterpreted in this context.
For the \WyyShort analysis, the full reconstruction level template was generated for all samples, and is used for reinterpretation.
In the initial publication of \ZvvyShort and \WZShort, which used reco-level results, only a subset of operators was probed, so a template replacement procedure is developed to derive the kinematic shapes of the missing operators.
In this procedure, the histogram template of the differential yields of the missing operator in the observable of interest is approximated using the reco-level template of an existing operator that has the lowest $\chi^2$ difference between corresponding templates at particle-level.
Differences between cross-sections are accounted for in the replacement procedure.
Dedicated non-closure studies on this procedure are conducted by applying the replacement in situations where the original templates exist. The difference in the confidence intervals obtained from the two templates is found to be at the percent level in all cases, except for the \ZvvyShort analysis that found an impact of 9\% on the intervals for the $f_{M1}$ and $f_{T9}$ coefficients.  A residual closure study is performed adding a conservative 30\% difference directly as an additional source of uncertainty in the \ZvvyShort analysis for all operators, with the corresponding confidence intervals widened by at most 4\%, and usually at the sub-percent level. Due to the negligible effect of the residual non-closure uncertainty on the replacement procedure, it is not included for the complete set of final results presented here.

Cross-terms were not considered in the original publications except for the \WZShort and \ssWWShort analyses in several specific cases. As such, cross-term effects at reco-level are simulated through the same replacement procedure.

% End of text imported from the .//tex/Inputs.tex input file

% The next lines are included from the .//tex/StatisticalModel.tex input file
\section{Statistical model}
\label{sec:stats}

The combination of the individual channels proceeds via a joint simultaneous likelihood model of the discriminating variables across all of the channels. The Wilson coefficients $f_{i}$ are treated as parameters of interest of the fit model. Systematic uncertainties are treated as nuisance parameters, denoted by $\bm\theta$. %
A likelihood ratio test statistic $\lambda$, is built from the likelihood to evaluate 68\% and 95\% confidence level intervals for the Wilson coefficients:
\begin{equation}
\lambda = -2\mathrm{log}\frac{L(\bm{\mu_i}, \hat{\hat{\bm{\theta}}})}{L(\hat{\bm{\mu_i}},\hat{\bm{\theta}})}
\end{equation}
where $\mu_i=f_i/\Lambda$, $L(\bm{\mu_i}, \hat{\hat{\bm{\theta}}})$ is the maximum of the likelihood for a given $\mu_{i}$, and $L(\hat{\bm{\mu_i}},\hat{\bm{\theta}})$ is the absolute maximum of the likelihood.
The total likelihood is the product of the individual likelihoods from the individual channels, which have not been changed relative to the original publications except for details mentioned earlier in this Letter.
For each given set of Wilson coefficients, a maximum-likelihood fit is performed by setting other coefficients to zero and maximizing the likelihood relative to the nuisance parameters.
Confidence intervals are derived using Wilks' theorem, assuming that the profile likelihood test statistic is $\chi^2$-distributed~\cite{Cowan:2010js} with degrees of freedom matching the number of fitted Wilson coefficients.
The validity of Wilks' theorem, which holds asymptotically in the large sample limit, is tested against a full pseudo-experiment procedure in several cases, and the reported 68/95\% confidence level intervals are validated within 5\%.
Expected confidence intervals are computed from synthetic data constructed such that the estimators of all parameters match the true parameters of the SM-only hypothesis~\cite{Cowan:2010js}.

Since each analysis is statistically limited, the choice of correlation scheme between nuisance parameters of the input analysis has negligible impact on the final result, with the total confidence intervals changing by less than one percent in the extreme case where all systematic effects are removed.
None the less, all systematic uncertainties are treated as correlated across analyses if they originate from the same source. Thus all experimental uncertainties in the reconstruction and identification of final state jets, electrons, photons, muons, and missing transverse momentum, and the luminosity and pile-up uncertainties, are treated as fully correlated across all analyses.
There is an exception for jet-based systematic uncertainties in the \ZvvyShort analysis that are uncorrelated from the other analyses as a different jet definition is used.
Similarly heavy-flavor identification uncertainties are only correlated across analysis using the same algorithm definition and identification working point and uncorrelated otherwise.
Modeling uncertainties on the various background processes of each analysis are treated as uncorrelated due to the specific phase space of each analysis.
On the other hand, modeling uncertainties on the signal EFT processes, originating from QCD scale and parton distribution functions, are treated as correlated across all VBS analyses, with the \WyyShort process left uncorrelated due to the different initial parton contributions and relevant energy scales.
For the \ZvvyShort and \WyyShort analyses, all systematic uncertainties were re-evaluated using the finer discriminant binning explored in this work.

Several fits are performed for different interpretations of the data. One-dimensional fits are performed for each operator allowing only its matching Wilson coefficient to vary in the fit, while the remaining are set to zero.
Two-dimensional confidence intervals are presented for pairs of operators using a similar fit model where only the respective Wilson coefficients are non-zero and allowed to vary.
Profiled likelihood confidence intervals are also presented where all Wilson coefficients are allowed to float simultaneously. Profiled results enable the interpretation of the results with phenomenological models that predict simultaneous changes to multiple Wilson coefficients.

% End of text imported from the .//tex/StatisticalModel.tex input file

% The next lines are included from the .//tex/Results.tex input file
\section{Results}
\label{sec:results}

The 68\% and 95\% confidence level intervals for each Wilson coefficient are shown in Figure~\ref{fig:summary_unclip} (middle panel), which are obtained from fits where all but the relevant coefficient are fixed to zero.
The contribution of each input channel on the combined confidence intervals is shown in Figure \ref{fig:summary_unclip} (top-panel) and is calculated as in Equation~\ref{eq:impact}, where $\sigma_i$ is the 68\% confidence level intervals obtained in channel $i$ for a given Wilson coefficient:
\begin{equation}
\mathrm{Contribution(chan.)} = \frac{\frac{1}{\sigma^{2}_\mathrm{chan.}}}{\sum^{N_\mathrm{chan.}}_{i}\frac{1}{\sigma^{2}_{i}}}
\label{eq:impact}
\end{equation}
The sensitivity reach in terms of the new physics scale $\Lambda$ for each Wilson coefficient is shown at the 95\% confidence level intervals in Figure~\ref{fig:summary_unclip} (bottom-panel) and compared to bounds set by unitarity constraints.
The best individual constraint on the  $f_{S02}$, $f_{S1}$, $f_{M0}$, $f_{M1}$, and $f_{M7}$ Wilson coefficients originate from the \VVSemiShort analyses, while the best individual result on the $f_{M2}$, $f_{M3}$, $f_{M4}$, $f_{M5}$, and all tensor-type Wilson coefficients come from the \ZvvyShort channel.
The intervals on the scalar and mixed-type operators are tightened by up to 20\% in the combined fit relative to those derived using individual channels.
Due to the dominant sensitivity in the \ZvvyShort analysis, the tensor-type operator constraints are negligibly improved in the combined fit.

\begin{figure}[!t]
\centering
\includegraphics[width=.9\linewidth]{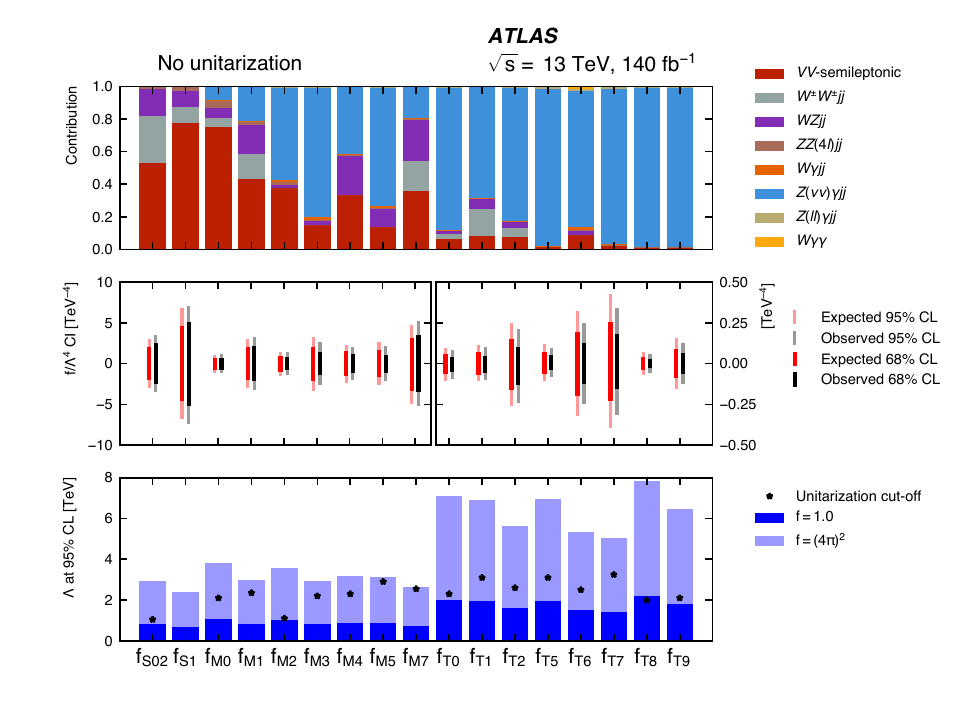}
\caption{Contributions of the individual channels to the combination (top), expected and observed 68\% and 95\% confidence level intervals from the combined fit (middle) and the probed energy scale for two illustrative values of the Wilson coefficients (bottom). No constraints from unitarity bounds are applied. The energy scale of the observed unitarization cut-off as evaluated through a clipping procedure is overlaid on the bottom panel. No unitarization cut-off is found for the $f_{S1}$ operator where the experimental interval is always weaker than the unitarity constraint.}
\label{fig:summary_unclip}
\end{figure}

As discussed in Section~\ref{sec:theory}, operators with mass dimension greater than four can lead to interaction probabilities that increase until unitarity is violated.
Following Ref~\cite{CMS-SMP-19-012}, the confidence intervals are recalculated for cases where the EFT contributions are zeroed for energy scales above a selected value, known as the clipping threshold.
This method provides an easily reproducible result, and provides a conservative limit as any deviations above the energy-scale are not considered consistent with the EFT\@.
The energy scale is given by the diboson invariant mass for VBS analyses and by the largest of the three possible diboson invariant masses for the tri-boson analysis. The confidence intervals are recalculated for clipping thresholds of 0.7, 1, 1.5, 2 and 3~\TeV, with the results for the 1.5~\TeV clipping threshold shown in Figure~\ref{fig:summary_clip1500}.
It can be seen that the combined intervals on the Wilson coefficients at the 1.5~\TeV clipping threshold are approximately an order of magnitude wider than the non-unitarized intervals.
The input channels most sensitive to the clipping threshold are the \VVSemiShort and \ZvvyShort  channels, where the fitted observables have a large reach in energy. Consequently, it can be seen at the 1.5 TeV clipping threshold that the contribution to the interval of some coefficients is more distributed across the input channels, with the \WZShort channel providing a similar contribution to previously-dominant \VVSemiShort and \ZvvyShort channels.

\begin{figure}[!t]
\centering
\includegraphics[width=.9\linewidth]{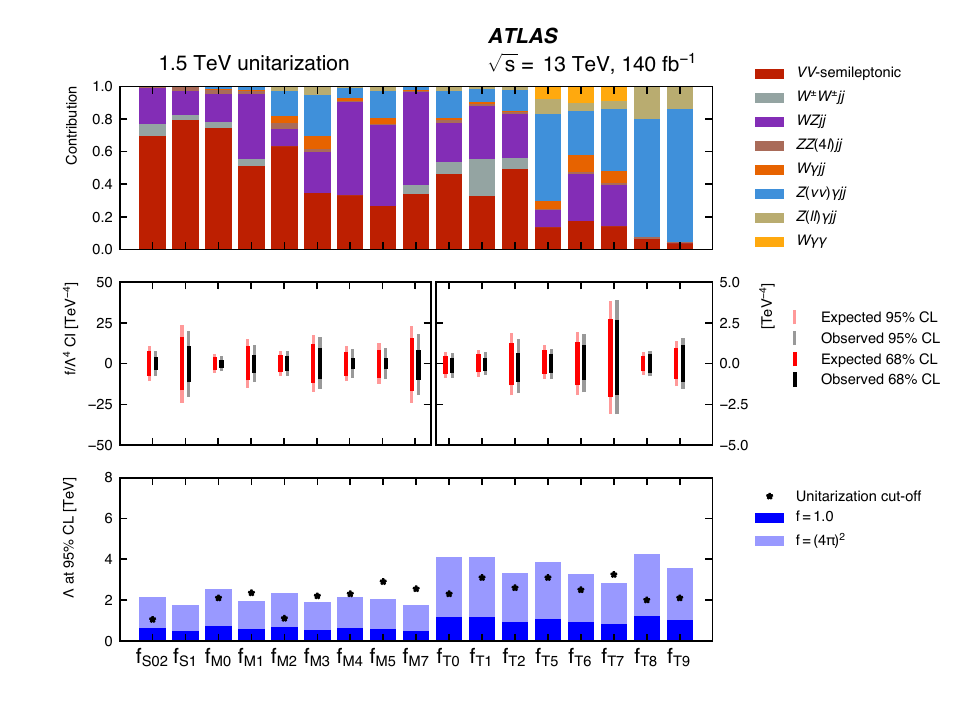}
\caption{Contributions of the individual channels to the combination (top), expected and observed 68\% and 95\% confidence level intervals from the combined fit (middle) and the probed energy scale for two illustrative values of the Wilson coefficients (bottom). Unitarity is preserved by setting higher dimension operator contributions to zero above a clipping threshold of 1.5~\TeV. The energy scale of the observed unitarization cut-off as evaluated through a clipping procedure is overlaid on the bottom panel. No unitarization cut-off is found for the $f_{S1}$ coefficient where the experimental interval is always weaker than the unitarity constraint.}
\label{fig:summary_clip1500}
\end{figure}

Unitarized confidence intervals are chosen as the intersection between the interpolated confidence intervals as a function of clipping threshold and the theoretical unitarity bounds provided in Ref~\cite{eboli2020,eboli2026unitarity}.
The clipping threshold at the unitarized point is designated as the unitarization cut-off.
Examples of the confidence intervals scans across clipping thresholds for the combined fit on the $f_{S02}$, $f_{M0}$, and $f_{T0}$ Wilson coefficients are shown in Figure~\ref{fig:clipping_scans}.
Where applicable, positivity bounds from Ref~\cite{positivity_bounds} are shown.
A unitarization cut-off is found in all cases except for the $f_{S1}$ coefficient, where the theory bound is tighter than the derived experimental interval at all clipping thresholds.

\begin{figure}[!t]
\centering
\subfloat[]{\includegraphics[width=.45\linewidth]{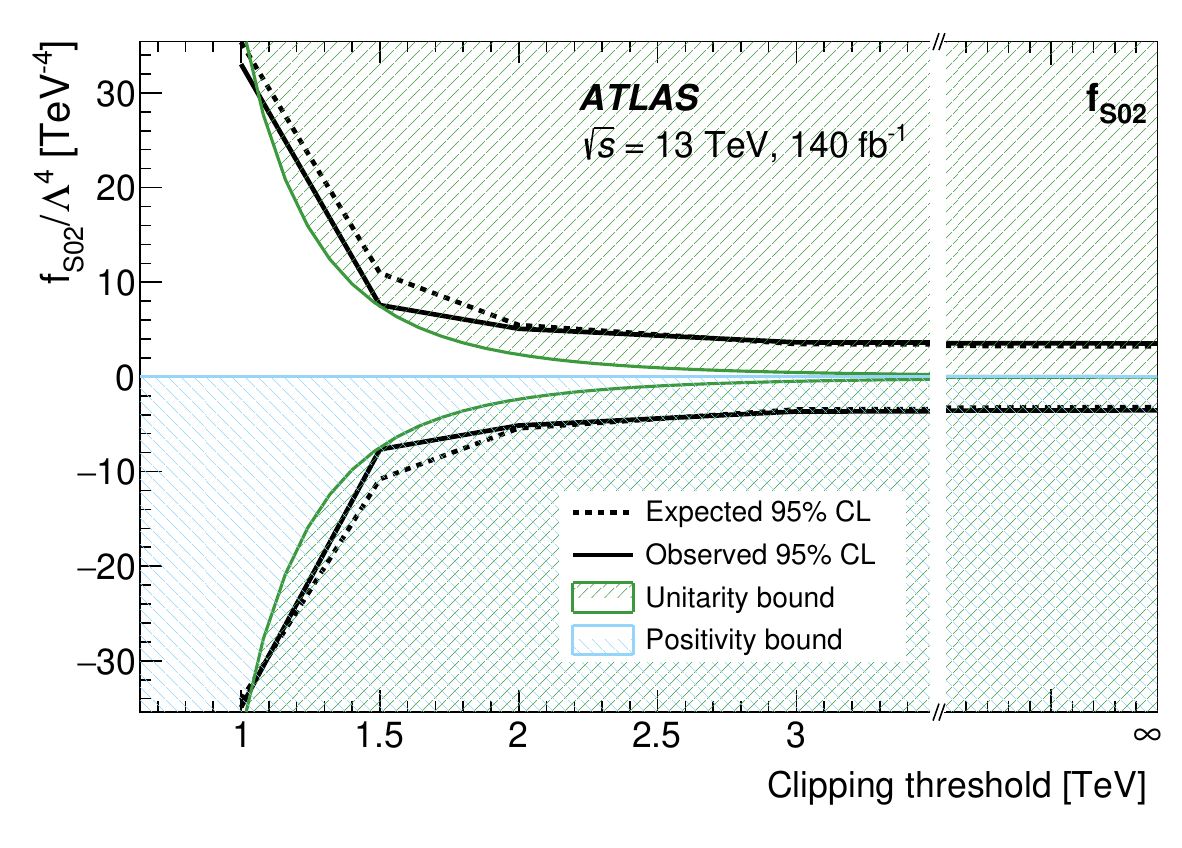}}
\subfloat[]{\includegraphics[width=.45\linewidth]{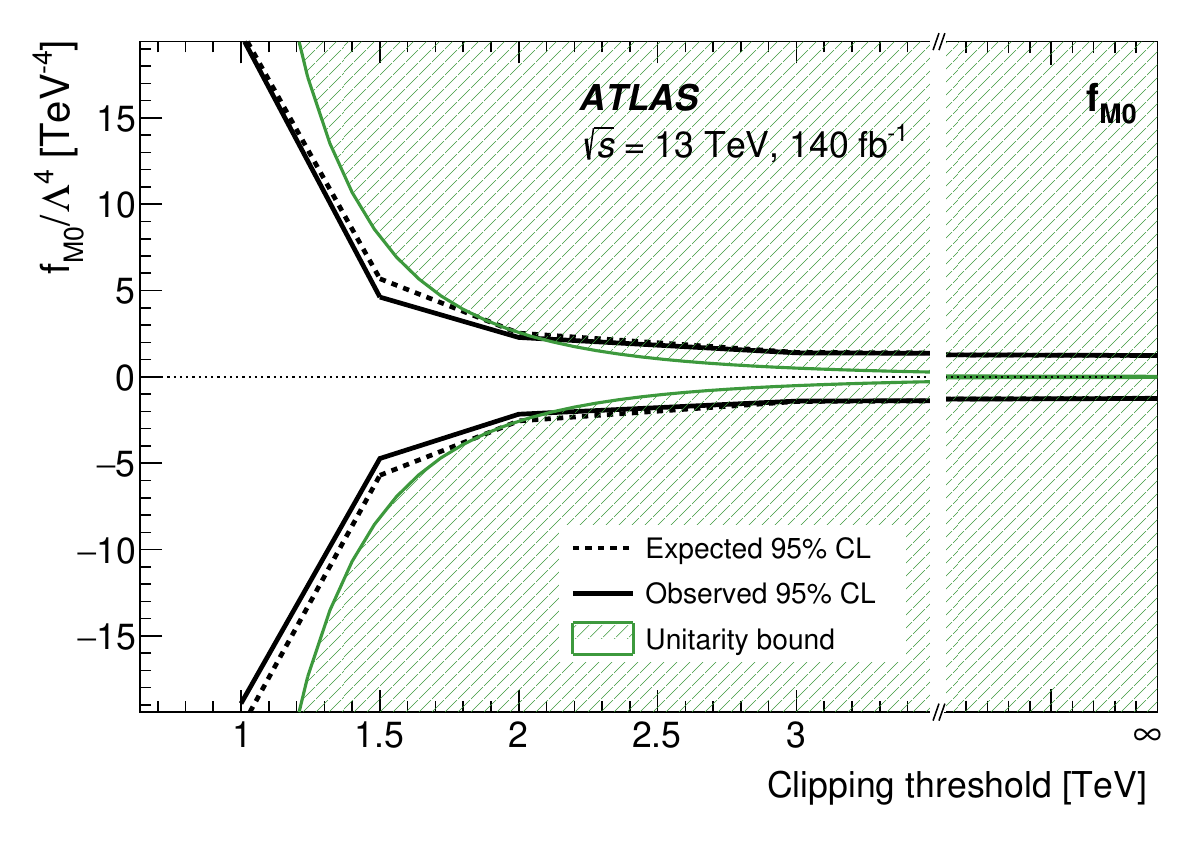}}\\
\subfloat[]{\includegraphics[width=.45\linewidth]{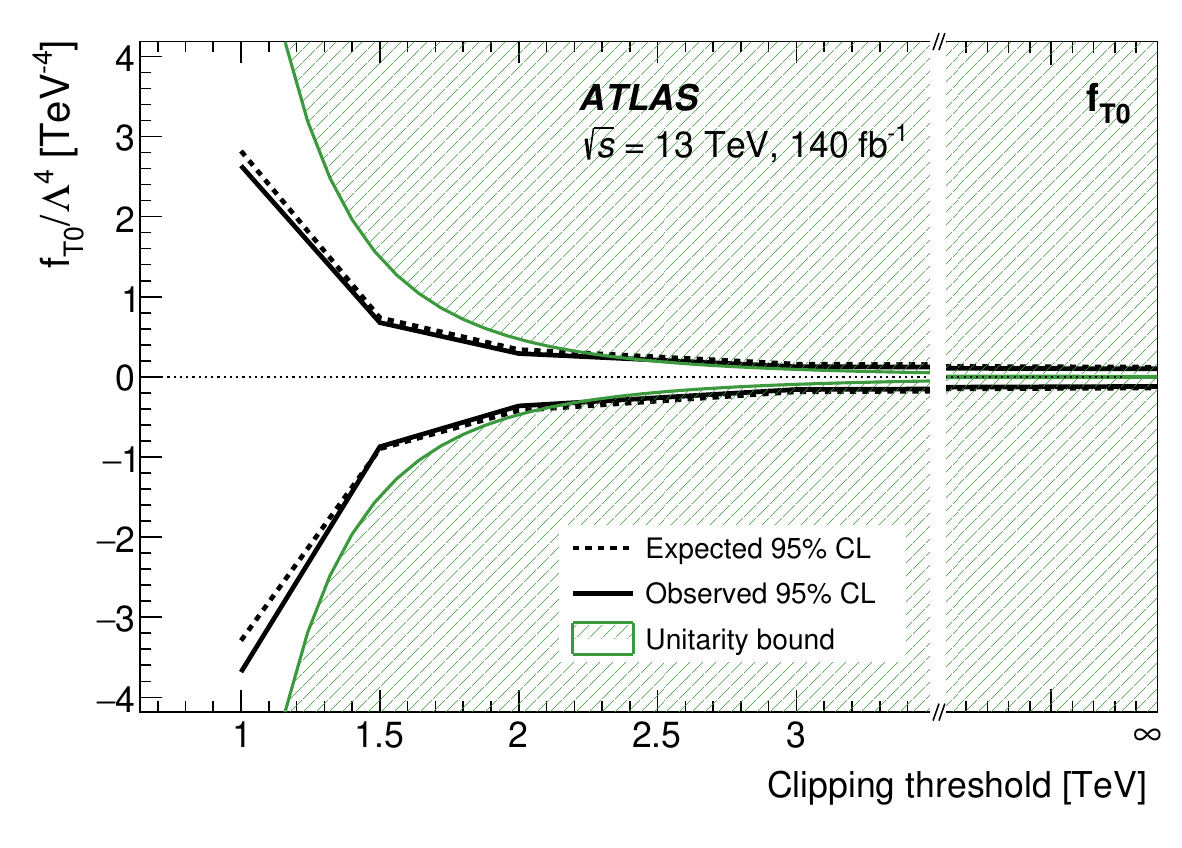}}
\caption{Expected and observed 95\% confidence level confidence intervals on the a) $f_{S02}$, b) $f_{M0}$ and c) $f_{T0}$ Wilson coefficients as a function of clipping threshold. Unitarity and positivity bounds are shown in green and blue respectively. Unitarized intervals are taken to be the intersection between the interpolated confidence intervals and the unitarity bound.}
\label{fig:clipping_scans}
\end{figure}

\begin{table}[!t]
\centering
\caption{Expected and observed 95\% confidence level intervals from the combined fit within the constraints of unitarity for each Wilson coefficient considered. Unitarity is preserved using the clipping procedure where the abscissa and ordinate of the intersection point of the unitarity bound and the confidence intervals as a function of clipping threshold is taken as the unitarization cut-off and interval respectively. No unitarized confidence interval is found for the $f_{S1}$ coefficient where the experimental interval is always weaker than the unitarity constraint.}
\resizebox{\textwidth}{!}{\begin{tabular}{ccccc}
\toprule
\multirow{2}{*}{Coefficient} & \multicolumn{2}{c}{Expected} & \multicolumn{2}{c}{Observed} \\
\cmidrule{2-5}
& Confidence interval [\TeV$^{-4}$] & Unitarization cut-off [\TeV] & Confidence Intervals [\TeV$^{-4}$] & Unitarization cut-off [\TeV]\\
\midrule
$f_{S02}/\Lambda^4$ & [-32.3, 34.2] & 1.0 & [-33.2, 30.1] & 1.0, 1.1\\
$f_{S1}/\Lambda^4$ & -- & -- & -- & -- \\
$f_{M0}/\Lambda^4$ & [-2.55, 2.52] & 2.0 & [-2.08, 2.22] & 2.1\\
$f_{M1}/\Lambda^4$ & [-5.72, 5.78] & 2.3 & [-5.44, 5.18] & 2.3, 2.4\\
$f_{M2}/\Lambda^4$ & [-15.8, 15.7] & 1.2 & [-23.7, 22.8] & 1.1\\
$f_{M3}/\Lambda^4$ & [-8.49, 8.49] & 2.0 & [-6.56, 6.63] & 2.2\\
$f_{M4}/\Lambda^4$ & [-4.78, 4.76] & 2.1 & [-3.59, 3.63] & 2.3\\
$f_{M5}/\Lambda^4$ & [-4.79, 4.79] & 2.5 & [-3.04, 3.10] & 2.9\\
$f_{M7}/\Lambda^4$ & [-8.50, 8.04] & 2.5 & [-7.90, 7.65] & 2.5, 2.6\\
$f_{T0}/\Lambda^4$ & [-0.40, 0.30] & 2.1, 2.3 & [-0.32, 0.22] & 2.2, 2.4\\
$f_{T1}/\Lambda^4$ & [-0.18, 0.18] & 3.0 & [-0.16, 0.16] & 3.1\\
$f_{T2}/\Lambda^4$ & [-0.71, 0.67] & 2.4 & [-0.61, 0.42] & 2.5, 2.7\\
$f_{T5}/\Lambda^4$ & [-0.26, 0.39] & 2.5, 2.7 & [-0.15, 0.17] & 3.0, 3.2\\
$f_{T6}/\Lambda^4$ & [-0.87, 0.85] & 2.2 & [-0.55, 0.57] & 2.5\\
$f_{T7}/\Lambda^4$ & [-0.69, 1.05] & 2.7, 3.0 & [-0.52, 0.56] & 3.2, 3.3\\
$f_{T8}/\Lambda^4$ & [-0.50, 0.50] & 1.8 & [-0.28, 0.28] & 2.0\\
$f_{T9}/\Lambda^4$ & [-0.65, 0.65] & 2.0 & [-0.53, 0.53] & 2.1\\
\bottomrule
\end{tabular}}
\label{tab:unitarized_limits}
\end{table}

The combined expected and observed unitarized confidence intervals, along with the energy threshold that maintains unitarity, are displayed in Table~\ref{tab:unitarized_limits}.
These combined results improve upon the previous published confidence intervals on these Wilson coefficients by 17--96\% depending on the coefficient.
The improvements in the unitarized results are larger than for the results without unitarity, due to steep trend in the theoretical unitarity bounds as a function of clipping threshold seen in Figure~\ref{fig:clipping_scans}, which favors sensitivity at low clipping threshold. In this regime more analyses have a non-negligible contribution to the combined results, which illustrates the merit of such a combination.
The confidence intervals are competitive with existing results from CMS~\cite{CMS-SMP-22-011,CMS-SMP-20-016,CMS-SMP-20-001,CMS-SMP-21-014,CMS-SMP-19-013,CMS-SMP-19-012}.

\begin{figure}[!t]
\centering
\includegraphics[width=.9\linewidth]{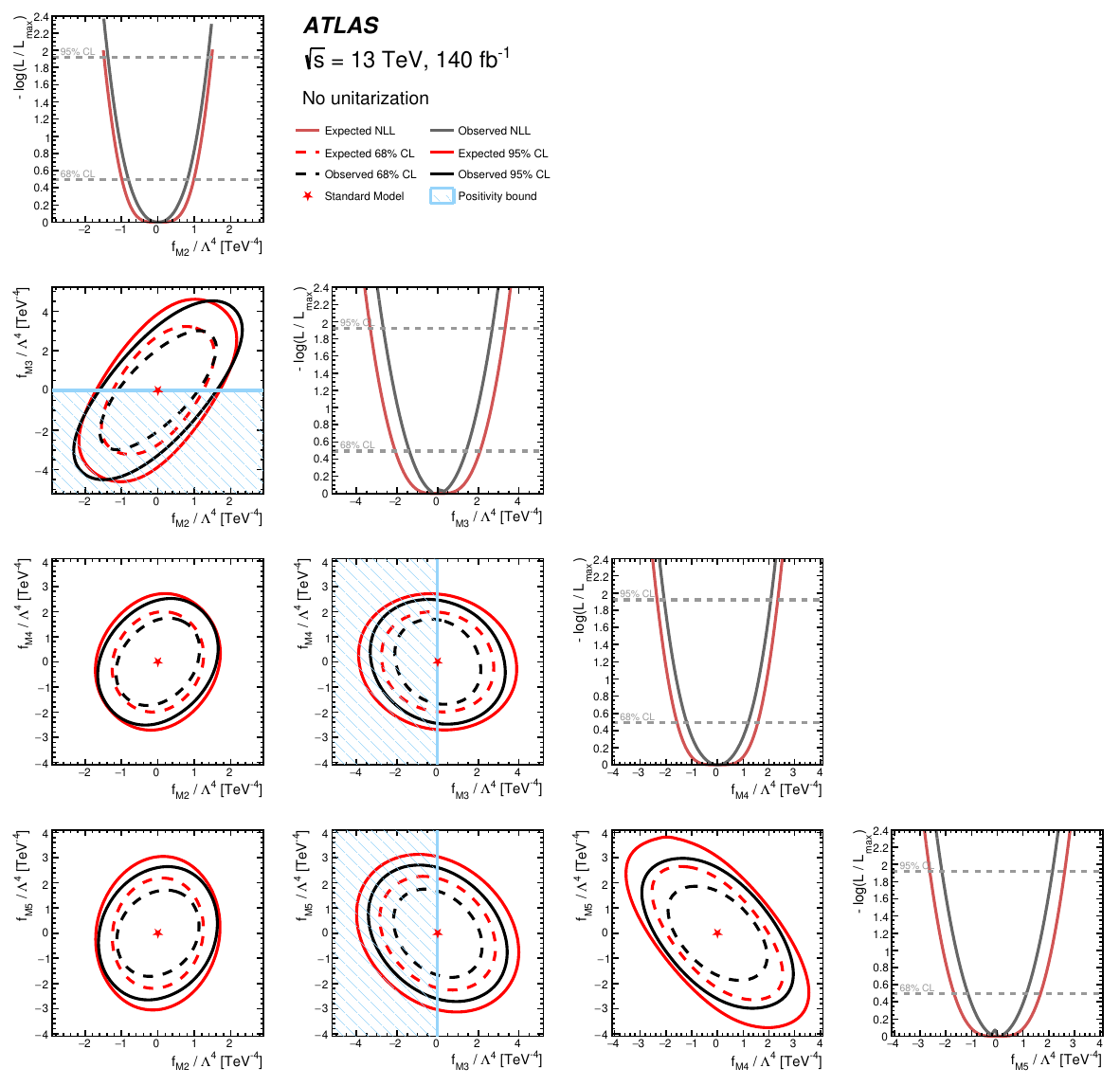}
\caption{Expected (red) and observed (black) 68\% (dashed) and 95\% (solid) confidence level contours when considering two of the $f_{M2}$, $f_{M3}$, $f_{M4}$, and $f_{M5}$ coefficients simultaneously alongside the relevant cross-term. The expected (greyed-red) and observed (grey) negative log-likelihood (NLL) curve for each of the operators when considering just one coefficient at a time is also shown. Unitarity constraints are not visible and positivity bounds are overlaid as a blue filled region.}
\label{fig:2Dlimits}
\end{figure}

Results were investigated for pairs of Wilson coefficients that have an appreciable cross-term, while keeping the others set to zero.
The size of cross-terms is highly correlated with the structure of the \Eboli operator basis, with high-suppression between scalar, mixed, and tensor type families, and block-structure within each.
For instance, the operators associated with the $f_{M0}$, $f_{M1}$, and $f_{M7}$ coefficients form a mostly orthogonal subset relative to the $f_{M2}$, $f_{M3}$, $f_{M4}$ and $f_{M5}$ in the mixed-type family.
Two dimension exclusion contours at 68\% and 95\% confidence level for the $f_{M2}$, $f_{M3}$, $f_{M4}$, and $f_{M5}$ pairings can be seen in Figure~\ref{fig:2Dlimits}, without clipping, alongside with the individual negative-log likelihood scans projections along each coefficient direction in the fit. Two dimensional results are obtained for each clipping threshold.
Positivity bounds are shown where applicable, and reduce the constrained phase space drastically in such cases.

\begin{table}[!t]
\caption{Expected and observed 95\% confidence level intervals allowing all Wilson coefficients to be fitted simultaneously}
\centering
\begin{tabular}{ccc}
\toprule
Coefficient & Profiled expected confidence intervals [\TeV$^{-4}$] & Profiled observed confidence intervals [\TeV$^{-4}$]\\
\midrule
$f_{S02}/\Lambda^4$& [-4.06, 4.08] & [-4.87, 4.88] \\
$f_{S1}/\Lambda^4$ & [-8.72, 8.73] & [-9.82, 9.82] \\
$f_{M0}/\Lambda^4$ & [-1.67, 1.68] & [-1.68, 1.65] \\
$f_{M1}/\Lambda^4$ & [-8.13, 7.52] & [-9.05, 8.58] \\
$f_{M2}/\Lambda^4$ & [-1.91, 1.91] & [-1.97, 1.99] \\
$f_{M3}/\Lambda^4$ & [-4.16, 4.16] & [-3.89, 3.92] \\
$f_{M4}/\Lambda^4$ & [-3.15, 3.15] & [-2.55, 2.56] \\
$f_{M5}/\Lambda^4$ & [-3.48, 3.49] & [-2.77, 2.77] \\
$f_{M7}/\Lambda^4$ & [-16.6, 15.7] & [-12.4, 12.9] \\
$f_{T0}/\Lambda^4$ & [-0.33, 0.24] & [-0.34, 0.22] \\
$f_{T1}/\Lambda^4$ & [-0.199, 0.211] & [-0.254, 0.283] \\
$f_{T2}/\Lambda^4$ & [-0.578, 0.639] & [-0.665, 0.599] \\
$f_{T5}/\Lambda^4$ & [-0.405, 0.384] & [-0.319, 0.289] \\
$f_{T6}/\Lambda^4$ & [-0.658, 0.642] & [-0.490, 0.370] \\
$f_{T7}/\Lambda^4$ & [-1.51, 1.58]   & [-1.17, 1.13] \\
$f_{T8}/\Lambda^4$ & [-0.196, 0.181] & [-0.163, 0.147] \\
$f_{T9}/\Lambda^4$ & [-0.360, 0.384] & [-0.281, 0.280] \\
\bottomrule
\end{tabular}
\label{tab:profiled_limits}
\end{table}

Fits are also performed allowing all Wilson coefficients to float in the fit.
Cross-term contributions are relevant in this fit, which, as explained above, are highly suppressed between operators of different scalar/mixed/tensor type, but can be sizeable within the types.
The expected and observed 95\% confidence level intervals from these profiled fits with no clipping applied are shown in Table~\ref{tab:profiled_limits}.
For coefficients without sizeable cross-term pairs, the profiled results differ by only a few percent relative to those obtained with the individual Wilson coefficient fits presented earlier.
More significant changes can be seen for $f_{M7}$, $f_{T0}$, $f_{T1}$, $f_{T5}$, $f_{T6}$, $f_{T7}$, $f_{T8}$, and $f_{T9}$ that have non-negligible cross terms with other operators.

% End of text imported from the .//tex/Results.tex input file

\FloatBarrier

% The next lines are included from the .//tex/Conclusion.tex input file
\section{Conclusions}
\label{sec:conclusion}

This Letter details a combination of VBS and tri-boson measurements performed using $pp$ collision data collected by the ATLAS detector at $\sqrt{s} = 13$~\TeV at the LHC to optimally constrain dimension-8 effective field theory operators producing \aQGCs. This is the most comprehensive set of aQGC EFT results to date and the confidence intervals are competitive with similar results from CMS.

Results are presented in the form of 68\% and 95\% confidence level intervals, from fits to single Wilson coefficients with others set to zero, and from two dimensional fits for operator pairs with non-negligible cross terms.
It is found that the 68\% and 95\% confidence level intervals on the coefficients without unitarization applied are mainly driven by the \Zvvy and \VVSemi channels, depending on the operator, with the combination bringing up to a 20\% improvement on individual results.
Results respecting unitarity constraints are presented following the clipping method.
At a clipped energy threshold of 1.5~\TeV the relative contributions of each input analysis are more even, and the combined bounds are found to improve the published results by up to 96\%.
Confidence intervals are also presented from profiled fits where all Wilson coefficients are free to vary simultaneously.

% End of text imported from the .//tex/Conclusion.tex input file

%

%
%
%
%
%
\section*{Acknowledgements}
%

%

% The next lines are included from the .//acknowledgements/Acknowledgements.tex input file
%
%

%
%

We thank CERN for the very successful operation of the LHC and its injectors, as well as the support staff at
CERN and at our institutions worldwide without whom ATLAS could not be operated efficiently.

The crucial computing support from all WLCG partners is acknowledged gratefully, in particular from CERN, the ATLAS Tier-1 facilities at TRIUMF/SFU (Canada), NDGF (Denmark, Norway, Sweden), CC-IN2P3 (France), KIT/GridKA (Germany), INFN-CNAF (Italy), NL-T1 (Netherlands), PIC (Spain), RAL (UK) and BNL (USA), the Tier-2 facilities worldwide and large non-WLCG resource providers. Major contributors of computing resources are listed in Ref.~\cite{ATL-SOFT-PUB-2026-001}.

We gratefully acknowledge the support of ANPCyT, Argentina; YerPhI, Armenia; ARC, Australia; BMWFW and FWF, Austria; ANAS, Azerbaijan; CNPq and FAPESP, Brazil; NSERC, NRC and CFI, Canada; CERN; ANID, Chile; CAS, MOST and NSFC, China; Minciencias, Colombia; MEYS CR, Czech Republic; DNRF and DNSRC, Denmark; IN2P3-CNRS and CEA-DRF/IRFU, France; SRNSFG, Georgia; BMFTR, HGF and MPG, Germany; GSRI, Greece; RGC and Hong Kong SAR, China; ICHEP and Academy of Sciences and Humanities, Israel; INFN, Italy; MEXT and JSPS, Japan; CNRST, Morocco; NWO, Netherlands; RCN, Norway; MNiSW, Poland; FCT, Portugal; MNE/IFA, Romania; MSTDI, Serbia; MSSR, Slovakia; ARIS and MVZI, Slovenia; DSI/NRF, South Africa; MICIU/AEI, Spain; SRC and Wallenberg Foundation, Sweden; SERI, SNSF and Cantons of Bern and Geneva, Switzerland; NSTC, Taipei; TENMAK, T\"urkiye; STFC/UKRI, United Kingdom; DOE and NSF, United States of America.

Individual groups and members have received support from BCKDF, CANARIE, CRC and DRAC, Canada; CERN-CZ, FORTE and PRIMUS, Czech Republic; COST, ERC, ERDF, Horizon 2020 and Marie Sk{\l}odowska-Curie Actions, European Union; Investissements d'Avenir Labex, Investissements d'Avenir Idex and ANR, France; DFG and AvH Foundation, Germany; Herakleitos, Thales and Aristeia programmes co-financed by EU-ESF and the Greek NSRF, Greece; BSF-NSF and MINERVA, Israel; NCN and NAWA, Poland; La Caixa Banking Foundation, CERCA and AGAUR programs from Generalitat de Catalunya and PROMETEO and GenT Programmes Generalitat Valenciana, Spain; G\"{o}ran Gustafssons Stiftelse, Sweden; The Royal Society and Leverhulme Trust, United Kingdom; Eric and Wendy Schmidt Fund for Strategic Innovation, United States of America.

In addition, individual members wish to acknowledge support from Chile: Agencia Nacional de Investigaci\'on y Desarrollo (ANID FONDECYT reg. 1230987, FONDECYT 1230812, FONDECYT 1240864, Fondecyt 3240661, Fondecyt Regular 1240721); China: Chinese Ministry of Science and Technology (MOST-2023YFA1605700, MOST-2023YFA1609300), National Natural Science Foundation of China (NSFC 12275265, NSFC-W2543005); Czech Republic: Czech Science Foundation (GACR - 24-11373S), Ministry of Education Youth and Sports (ERC-CZ-LL2327, FORTE CZ.02.01.01/00/22\_008/0004632), PRIMUS Research Programme (PRIMUS/21/SCI/017); EU: H2020 European Research Council (ERC - 101002463); European Union: European Research Council (BARD No. 101116429, ERC - 101219398, ERC - 948254, ERC 101089007), European Regional Development Fund (HE COFUND GA No.101081355, ERDF), Marie Sklodowska-Curie Actions (GAP-101168829); France: Agence Nationale de la Recherche (ANR-21-CE31-0013, ANR-22-EDIR-0002, ANR-24-CE31-0504-01); Germany: Deutsche Forschungsgemeinschaft (DFG - 469666862); China: Research Grants Council (GRF); Italy: Istituto Nazionale di Fisica Nucleare (LHC-MIUR - 28003/2025), Ministero dell'Università e della Ricerca (NextGenEU 153D23001490006 M4C2.1.1, NextGenEU I53D23000820006 M4C2.1.1, NextGenEU I53D23001490006 M4C2.1.1, SOE2024\_0000023); Japan: Japan Society for the Promotion of Science (JSPS KAKENHI  JP25H0063, JSPS KAKENHI JP22H01227, JSPS KAKENHI JP22H04944, JSPS KAKENHI JP22KK0227, JSPS KAKENHI JP24K23939, JSPS KAKENHI JP24KK0251, JSPS KAKENHI JP25H00650, JSPS KAKENHI JP25H01291, JSPS KAKENHI JP25K01011, JSPS KAKENHI JP25K01023); Poland: Polish National Science Centre (NCN 2021/42/E/ST2/00350, NCN OPUS 2023/51/B/ST2/02507, NCN OPUS nr 2022/47/B/ST2/03059, NCN UMO-2019/34/E/ST2/00393, UMO-2022/47/O/ST2/00148, UMO-2023/49/B/ST2/04085, UMO-2023/51/B/ST2/00920, UMO-2024/53/N/ST2/00869); Spain: Agència de Gestió d'Ajuts Universitaris i de Recerca. (AGAUR - 2023 BP 00141), Ministry of Science and Innovation (RYC2019-028510-I, RYC2020-030254-I, RYC2021-031273-I, RYC2022-038164-I), Ministerio de Ciencia, Innovación y Universidades/Agencia Estatal de Investigaci\'on (EU NextGenerationEU (PRTR-C17.I1), PID2022-142604OB-C22); Sweden: Carl Trygger Foundation (Carl Trygger Foundation CTS 22:2312), Swedish Research Council (Swedish Research Council 2023-04654, VR 2021-03651, VR 2022-03845, VR 2022-04683, VR 2023-03403, VR 2024-05451, VR 2025-05940), Knut and Alice Wallenberg Foundation (KAW 2023.0366); Switzerland: Swiss National Science Foundation (SNSF - PCEFP2\_194658); United Kingdom: The Binks Trust, Royal Society (NIF-R1-231091); United States of America: U.S. Department of Energy (ECA DE-AC02-76SF00515), John Templeton Foundation (John Templeton Foundation 63206), Neubauer Family Foundation.

%
%

% End of text imported from the .//acknowledgements/Acknowledgements.tex input file

%
%
%
%
%
%

%
%
\printbibliography
\clearpage
% ATLAS Collaboration author list
% Reference date of STDM-2023-21 is 2026-02-16
% Author list last updated on date 16-JUN-26
% Data extracted on 16-Jun-2026 for paper reference STDM-2023-21
% at 5:00pm
 
\begin{flushleft}
\hypersetup{urlcolor=black}
{\Large The ATLAS Collaboration}

\bigskip

\AtlasOrcid[0000-0002-6665-4934]{G.~Aad}$^\textrm{\scriptsize 102}$,
\AtlasOrcid[0000-0001-7616-1554]{E.~Aakvaag}$^\textrm{\scriptsize 17}$,
\AtlasOrcid[0000-0002-5888-2734]{B.~Abbott}$^\textrm{\scriptsize 121}$,
\AtlasOrcid[0000-0002-0287-5869]{S.~Abdelhameed}$^\textrm{\scriptsize 83b}$,
\AtlasOrcid[0000-0002-1002-1652]{K.~Abeling}$^\textrm{\scriptsize 54}$,
\AtlasOrcid[0000-0001-5763-2760]{N.J.~Abicht}$^\textrm{\scriptsize 48}$,
\AtlasOrcid[0000-0002-8496-9294]{S.H.~Abidi}$^\textrm{\scriptsize 30}$,
\AtlasOrcid[0009-0003-6578-220X]{M.~Aboelela}$^\textrm{\scriptsize 44}$,
\AtlasOrcid[0000-0002-9987-2292]{A.~Aboulhorma}$^\textrm{\scriptsize 36e}$,
\AtlasOrcid[0000-0001-5329-6640]{H.~Abramowicz}$^\textrm{\scriptsize 154}$
% ...
\bigskip
\\

% ...
$^{17}$Department for Physics and Technology, University of Bergen, Bergen; Norway.\\
% ...
$^{30}$Physics Department, Brookhaven National Laboratory, Upton NY; United States of America.\\
% ...
$^{36}$$^{(a)}$Facult\'e des Sciences Ain Chock, Universit\'e Hassan II de Casablanca;$^{(b)}$Facult\'{e} des Sciences, Universit\'{e} Ibn-Tofail, K\'{e}nitra;$^{(c)}$Facult\'e des Sciences Semlalia, Universit\'e Cadi Ayyad, LPHEA-Marrakech;$^{(d)}$LPMR, Facult\'e des Sciences, Universit\'e Mohamed Premier, Oujda;$^{(e)}$Facult\'e des sciences, Universit\'e Mohammed V, Rabat;$^{(f)}$Institute of Applied Physics, Mohammed VI Polytechnic University, Ben Guerir; Morocco.\\
% ...
$^{44}$Physics Department, Southern Methodist University, Dallas TX; United States of America.\\
% ...
$^{48}$Fakult\"{a}t Physik, Technische Universit{\"a}t Dortmund, Dortmund; Germany.\\
% ...
$^{54}$II. Physikalisches Institut, Georg-August-Universit\"{a}t G\"ottingen, G\"ottingen; Germany.\\
% ...
$^{83}$$^{(a)}$Khalifa University of Science and Technology, Abu Dhabi;$^{(b)}$New York University Abu Dhabi, Abu Dhabi;$^{(c)}$United Arab Emirates University, Al Ain;$^{(d)}$University of Sharjah, Sharjah; United Arab Emirates.\\
% ...
$^{102}$CPPM, Aix-Marseille Universit\'e, CNRS/IN2P3, Marseille; France.\\
% ...
$^{121}$Homer L. Dodge Department of Physics and Astronomy, University of Oklahoma, Norman OK; United States of America.\\
% ...
$^{154}$Raymond and Beverly Sackler School of Physics and Astronomy, Tel Aviv University, Tel Aviv; Israel.\\
% ...
$^{a}$ Also at Affiliated with an institute formerly covered by a cooperation agreement with CERN.\\
$^{b}$ Also at An-Najah National University, Nablus; Palestine.\\
$^{c}$ Also at Borough of Manhattan Community College, City University of New York, New York NY; United States of America.\\
$^{d}$ Also at Center for Interdisciplinary Research and Innovation (CIRI-AUTH), Thessaloniki; Greece.\\
$^{e}$ Also at Centre of Physics of the Universities of Minho and Porto (CF-UM-UP); Portugal.\\
$^{f}$ Also at CERN, Geneva; Switzerland.\\
% ...

\end{flushleft}

% Created with Glance <Atlas.Glance@cern.ch>

%

%
%

%

%
%
%
%

\end{document}